# DOTIFS: a new multi-IFU optical spectrograph for the 3.6-m Devasthal optical telescope

Haeun Chung[a,b], A. N. Ramaprakash[c], Amitesh Omar[d], Swara Ravindranath[c], Sabyasachi Chattopadhyay[c], Chaitanya V. Rajarshi[c], Pravin Khodade[c]
[a]Dept. of Physics and Astronomy, Seoul National Univ., Seoul, South Korea;
[b]Korea Institute for Advanced Study, Seoul, South Korea;
[c]Inter-University Centre for Astronomy & Astrophysics, Pune, India;
[d]Aryabhatta Research Institute of Observational Sciences, Nainital, India

## ABSTRACT

Devasthal Optical Telescope Integral Field Spectrograph (DOTIFS) is a new multi-object Integral Field Spectrograph (IFS) being designed and fabricated by the Inter-University Center for Astronomy and Astrophysics (IUCAA), Pune, India, for the Cassegrain side port of the 3.6m Devasthal Optical Telescope, (DOT) being constructed by the Aryabhatta Research Institute of Observational Sciences (ARIES), Nainital. It is mainly designed to study the physics and kinematics of the ionized gas, star formation and H II regions in the nearby galaxies. It is a novel instrument in terms of multi-IFU, built in deployment system, and high throughput. It consists of one magnifier, 16 integral field units (IFUs), and 8 spectrographs. Each IFU is comprised of a microlens array and optical fibers and has 7.4" × 8.7" field of view with 144 spaxel elements, each sampling 0.8" hexagonal aperture. The IFUs can be distributed on the telescope side port over an 8' diameter focal plane by the deployment system. Optical fibers deliver light from the IFUs to the spectrographs. Eight identical, all refractive, dedicated spectrographs will produce 2,304 R~1800 spectra over 370-740nm wavelength range with a single exposure. Volume Phase Holographic gratings are chosen to make smaller optics and get high throughput. The total throughput of the instrument including the telescope is predicted as 27.5% on average. Observing techniques, data simulator and reduction software are also under development. Currently, conceptual and baseline design review has been done. Some of the components have already been procured. The instrument is expected to see its first light in 2016.

## 1. INTRODUCTION

Integral Field Spectrograph (IFS) is an instrument which could obtain spectra of a two-dimensional object in one go. Thus, it can provide a large amount of spectral data in homogeneous condition, effectively. On account of its high multiplexing efficiency over traditional long-slit spectrographs, IFS instruments have become quite popular. Some examples of operating IFS instruments in various wavelength ranges are MUSE[1], PMAS[2], SAMI[3], VIMOS[4], and VIRUS-P[5]. (Table 1, specification summary) However, there is still a dearth of IFS instruments and DOTIFS is expected to help fill this demand. Moreover, the number of multi-IFU instruments is even fewer, which is to DOTIFS's great advantage. DOTIFS can observe up to 16 independent extend objects in a single exposure. The IFUs can be precisely and rapidly deployed by the built in automated deployment system. Eight high throughput dedicated spectrographs capture 2,304 spatial elements, at 0.8" spatial resolution, which is matched with the median natural seeing of the observing site. These capabilities are quite complementary to other IFS spectrographs. For example, MaNGA (PI Kevin Bundy), one of the next-generation instruments of the Sloan Digital Sky Survey, has much higher IFU multiplicity than the DOTIFS. However, it has relatively low spatial resolution because it is focused on survey functionality. Thus DOTIFS can be used as a follow up for MaNGA targets.

A few years ago, IUCAA developed a proto-type IFS, FIFUI[6], and commissioned it on the IUCAA 2m Girawali telescope. Subsequently, IUCAA proposed to build an advanced instrument for the upcoming (and India's largest) observatory at Devasthal in the Himalayan foothills at an altitude of about 2500m. The Devasthal Optical Telescope (DOT) is being set up by the Aryabhatta Research Institute of Observational Sciences (ARIES), Nainital, India.

Although the first proposal was made in 2011, it did not take off till the middle of 2012, until the Korean Institute of Advanced Study (KIAS) showed their interest in the project.

In this paper, we provide a brief overview of DOTIFS. We start with listing the key science drivers. In the following section, the instrument's specific requirements are addressed, along with its overall optical and mechanical structure. Finally, the current state of the data reduction and simulation software is shown, followed by the fabrication schedule and summary.

Table 1. Summary of IFS instruments specifications

| IFS Instrument | Telescope, Diameter | Spatial Sampling | Spectral Coverage | Spectral Resolution | Field of View, Field Shape | Multiplex Details |
|---|---|---|---|---|---|---|
| DOTIFS | DOT, 3.6m | 0.8" | 370-740 nm | 1200-2400 | 8', circular | 16 IFUs, 144 spaxels each |
| MaNGA | APO, 2.5m | 2" | 356-1040 nm | 1560-2650 | 3°, circular | 17 IFUs, 19–127 spaxels each |
| MUSE | VLT, 8.2m | 0.2", 0.025" | 465-930 nm | 2000-4000 | 1'x1', 7.5"x7.5", square | Single IFU, 90000 spaxels. |
| PMAS | CA, 3.5m | 2.68" | 350-1000 nm | 341-3404 | 74"x65", hexagonal | Single IFU, 331 spaxels |
| SAMI | AAT, 3.9m | 1.6" | 370-950 nm | 1300-8000 | 1°, circular | 13 IFUs, 61 spaxels each |
| VIMOS | VLT, 8.2m | 0.33", 0.67" | 370-1015 nm | 220-3100 | 13"x13", 54"x54", square | Single IFU, 6400 spaxels |
| VIRUS-P | HJST, 2.7m | 1.23" | 340-685 nm | 640-4600 | 1.7'x1.7', square | Single IFU, 246 spaxels |

## 2. SCIENCE DRIVERS

Following science topics have been proposed for DOTIFS. Each topic requires either wide IFS field capability or multi-IFU functionality. Here we only state the list of topics. Detail descriptions of each topic can be found in DOTIFS science reference document. (Publication in preparation)

- HII regions in our galaxy and nearby galaxies.
- Circum-nuclear rings in barred galaxies
- Merging, and Interacting galaxies, ultra-luminous infrared galaxies.
- Galaxies in clusters and Groups, Star formation as a function of environment
- Outer regions of star forming galaxies
- H$\alpha$ emission from Lyman-$\alpha$ clouds
- H$\alpha$ rotation curves of nearby galaxies
- AGN outflows, Dual AGNs or pairs
- Emission-line galaxies at intermediate redshifts

## 3. INSTRUMENT SPECIFICATIONS

The following instrument requirements flow down from the above science drivers.

- Wavelength coverage: 370 nm - 740nm.

    The shortest wavelength has been set by one of the major star formation indicators, OII doublet which is at a wavelength of 372.7nm in the rest frame. The longest wavelength is forced to be 740nm, since the DOTIFS spectrograph will employ only one dispersion element without any cross disperser. In this wavelength range, DOTIFS could detect OIII lines from z~0.4 galaxies.

- Spectral resolution: R ~ 1500-1800

    Many science drivers require the ability to measure velocity dispersions of galaxies. Observations of narrow line regions of the AGNs and measuring rotation velocity of the spiral galaxy can be made with R~1500-1800. It also enables to separation of Hβ and OIII lines in the blue region.

- Spatial resolution: 0.8 arcsec, vertex to vertex in hexagonal shape

    The spatial resolution of DOTIFS has been set to match with the median seeing of the observing site, which is about 0.8 arcsec. The tightly coupled relations between physical parameters of the input and output beams (f-ratio, beam size), microlens (size, f-ratio), and optical fiber (size, critical angle), also affects the realistically achievable spatial resolution. This spatial resolution resolves about 50 parsecs at a distance of 10 Mpc and this is sufficient to study regions of primary interests eg; NLR of AGNs, Giant HII regions, and starburst nuclear ring which lie in the range of hundreds to thousands parsecs.

- Field of view: 7.4" x 8.7" (IFU), 8' (deployable field)

    Since the typical sizes of the biggest luminous infrared galaxies at 100Mpc is about 40" x 40", IFU should be able to cover about 50" x 50" area. At the same time, to study the HII region structures in nearby galaxies or to observe multiple targets in a single exposure, multiple narrow field of view IFUs are preferred than a single wide field of view IFU. In addition, considering physical constraints of the spectrograph, one spectrograph could accommodate ~300 fibres. Given these considerations, each IFU is composed of 144 (12 x 12) spaxel elements, which correspond to 7.4" x 8.7" on the sky. Each spectrograph can then cover fibres from two independent IFUs. The size of the deployable area is determined by the field of view of the telescope side port, the physical size of the IFU deployable area and the optical performance of the magnifier. Considering all constraints, we get about 8' diameter circular field as the instrument field of view. This field is also comparable to the size of the optical disks of nearby galaxies.

- # of spatial elements: 2304

    The number of spatial elements is mostly constrained by the budget and the size of the available space at the location of spectrographs. Since we are able to accommodate up to eight spectrographs and one spectrograph can cover fibers from 2 IFUs, the number of spatial elements is determined as 2304.

- Throughput: 20%

    DOTIFS requires a flux sensitivity of $10^{-16}$ erg/cm$^2$/sec/angstrom/arcsec. To obtain a line S/N~5 from one hour exposure with the requirement, calculation suggests that the instrument (from the telescope to the spectrograph CCD) needs more than 20% total throughput.

Table 2 lists a summary of the above requirements.

Table 2. Main DOTIFS instrument specifications

| Specification | Requirements | Remarks |
|---|---|---|
| Wavelength coverage | 370nm – 740 nm | Covers entire optical range. |
| Spectral resolution | R ~ 1200 - 2400 | R~1200@ 370nm, R~2400@740nm |
| Spatial resolution | 0.8" | Vertex to vertex in hexagonal shape |
| Instrument field of view | 8' diameter | Limited by the magnifier performance |
| IFU Field of view | 7.4" x 8.7" | 12 x 12 microlenses |
| # of spaxel elements | 2304 | 16 IFUs |
| Throughput | 20% | From the telescope to the CCD |

## 4. INSTRUMENT DESIGNS

DOTIFS is mainly composed of four parts – the magnifier, IFUs, deployment system, and spectrographs. The magnifier enlarges the telescope's focal plane and modifies plate scale to match its image resolution with the physical size of the microlenses. IFUs gather lights from the focal plane and transfer them to the spectrograph via optical fibers. IFUs can be deployed on desired location on the focal plane by a deployment system. At this point of time, DOTIFS has passed the baseline design review and thus its initial design and parameters has been decided.

### 4.1 Magnifier

The plate scale at the Cassegrain focal plane of the 3.6m telescope is 1" per 157µm. To obtain a spatial resolution of 0.8", this translates to a microlens diameter of 126 µm. This does not match well with 100µm core,-125µm cladding diameter optical fibers. A fore-optics magnifier design with 5 singlets was used to modify (shown in Figure 1) the incoming f-number and magnify the telescope focal plane so as to obtain 0.8" per 300µm, with a corresponding f-ratio of 21.486 at the microlens surfaces. This optics also converts light from a calibration unit to mimic the beam from the telescope.

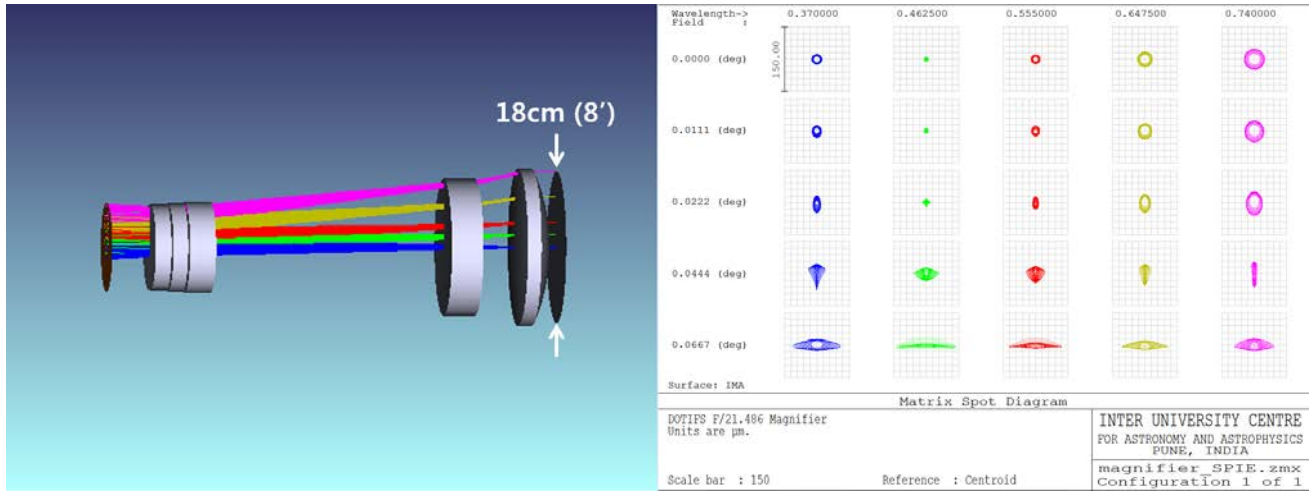

Figure 1. (Left) ZEMAX optical layout of the DOTIFS magnifier. The diameter of the image plane is 18cm, and it is correspond to 8' on the sky. (Right) Spot diagram of the magnifier. Note that the size of the scale box is 150µm (0.4") , and the width of the microlens is 300µm (0.8").

## 4.2 Integral Field Unit

As one of the most important parts of the DOTIFS, the IFUs capture light from an "area" of the sky and transfer them to the spectrograph. This is the reason why IFS is sometimes called as spectroscopic camera. Each IFU of DOTIFS is comprised of a 12x12 array of hexagonal shaped microlens and 144 optical fibers which are coupled with the microlens array. Fibers are attached at the back focus of the microlenses and deliver light from each lens to a spectrograph. The physical size of the hexagonal microlens is 300µm from vertex to vertex. The F/4.5 plano-convex microlenses, form pupils of diameter about 84µm at their back focus where fibres are in optical contact. The optical path inside the microlens is depicted in Figure 2. We are planning to use 100µm diameter, N.A. = 0.12 fiber from CeramOptec. Multiple fibre samples were tested in the lab for their input f-ratio as well as focal ration degradation FRD) effect.

## 4.3 Deployment System

The IFU deployment system of the DOTIFS is technically the most challenging part of the DOTIFS. It deploys IFUs at the exact position on the focal plane effectively, quickly, and safely without collision or other damage. We explored many possibilities to realize such a system and finally decided to utilize a one dimensional actuator and rail system. Current mechanical layout of the system is described in Figure 2. A linear actuator controls X-axis movement, and the rail system controls Y-axis movement. (z-axis is optical axis). A novel deployment algorithm is also required to be developed. It should allow observing techniques like dithering or nod-shuffle, too. Moreover, a careful collision evading system is also required in the deployment mechanism.

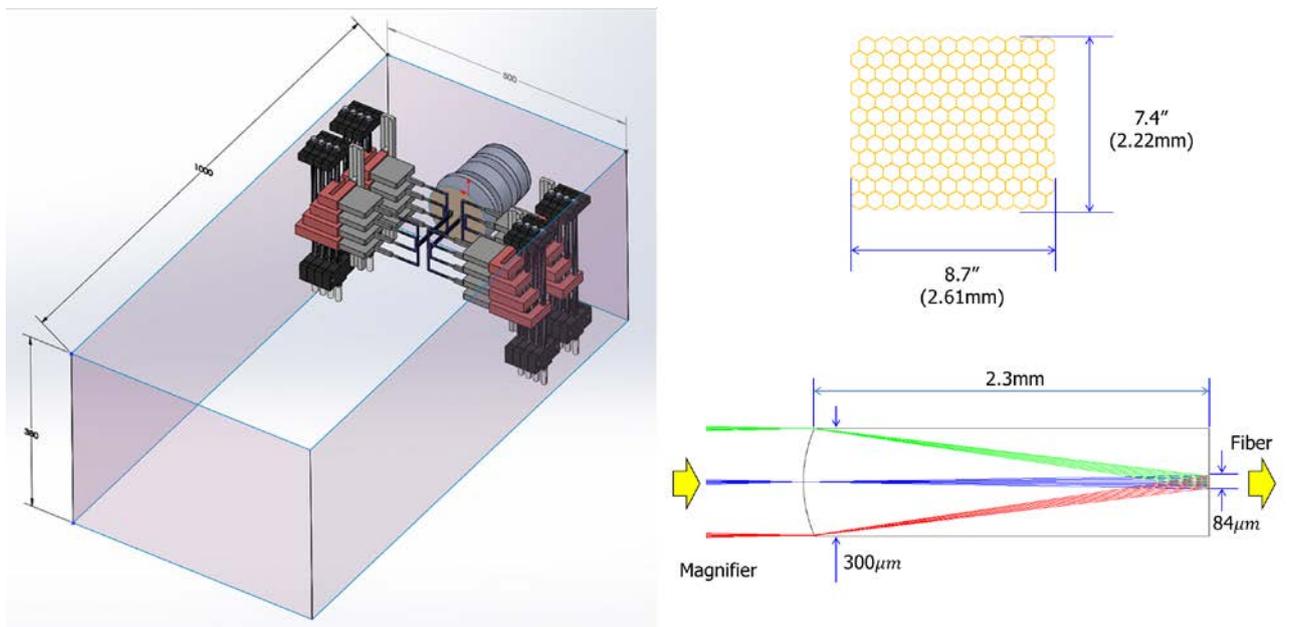

Figure 2. (Left) Mechanical layout of the deployment system. A rectangular box represents telescope's side port instrument box. (Top right) Schematic diagram of the IFU microlens array. (Bottom right) Path of the light inside the microlens. Lights arrive on IFUs are gathered and scrambled by microlens array and delivered to the spectrograph by optical fiber.

## 4.4 Spectrographs

Eight identical DOTIFS spectrographs deliver spectra from 2304 optical fibers with the desired spectral resolution. Each DOTIFS spectrograph handles 288 fibers coming from 2 IFUs. The spectrograph optics is composed of fiber slit, F/4 collimator, VPH grating, F/1.5 camera, and 4k by 2k CCD. (e2v CCD44-82, pixel size: 15µm) We use only 3k out of the 4k pixels in the spectral direction. Since the light coming out from the fiber is only a little slower than F/4, the spectrograph collimator must be at least F/4. Also, to qualify spectral resolution requirement, image size of the fiber at the CCD must be smaller than 37.5µm, which leads to a camera optics as fast as F/1.5. VPH grating and high throughput optics enables relatively high transmission spectrograph optics. The collimator is composed of 2 doublets and 3 singlets, and the camera is comprised of 3 doublets and 3 singlets, in total 16 lenses per spectrograph. The pupil size is 130mm in diameter. A broadband filter is positioned between the last element of the collimator and the pupil. The filter is inclined by 10 degrees to minimize multiple reflection ghosts. The VPH grating is located slightly behind the filter to be served as a dispersion element. It has 615 lines/mm, and slanted 8.49 degrees relative to the collimator optical axis to achieve desired spectral resolution. Figure 3 shows the optical layout and spot diagram of the spectrograph. Currently, vendor consultations are in progress for the spectrograph optics.

The predicted throughput performance of DOTIFS is shown in Figure 4. The shape of the throughput is mostly determined by the transmission of the VPH grating. The overall transmission is 27.5% on average, 34.2% at peak, 16.4% and 17.5% at the blue and red end, respectively. In principle, to maximize the grating throughput, the central wavelength should be same as the Littrow wavelength, which is where the Littrow ghost happens. There are several known ways to remove the ghost, but it may decrease the overall performance. Therefore, we decided to shift the ghost to a shorter wavelength (480nm) than the center (555nm), to minimize the effect of contamination by Littrow ghost instead of removing it. In this way, there will be a Littrow ghost at 480 nm, but since that wavelength is slightly shorter than major emission lines like Hβ or OIII lines, its influence on science is not significant.

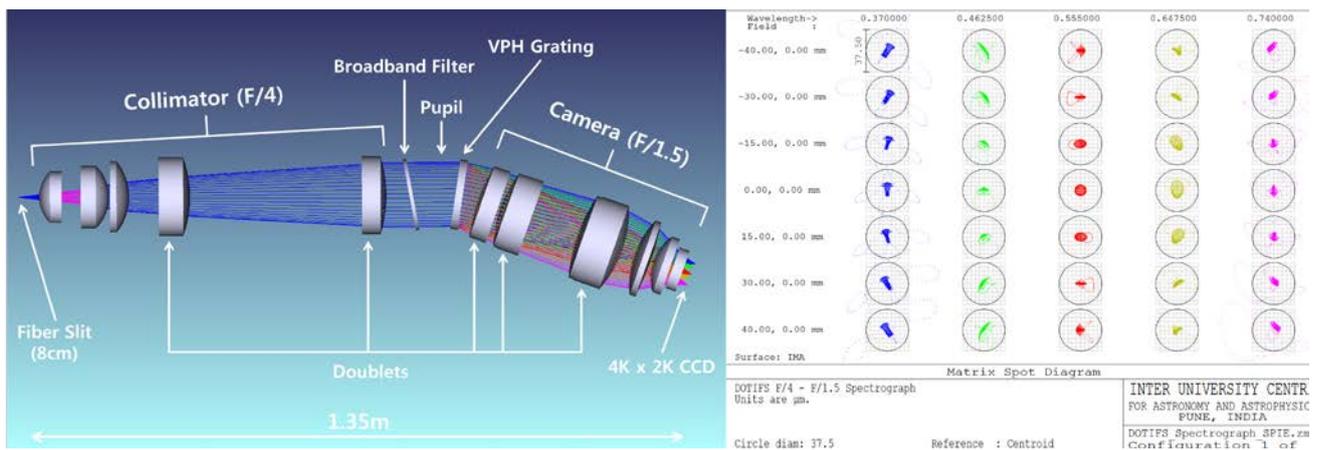

Figure 3. (Left) The Zemax optical layout of the spectrograph. The fiber slit is positioned along the direction perpendicular to the page. The broadband filter is inclined by an angle of 10 degrees to minimize the scattering. (Right) Matrix Spot diagram of the spectrograph. Note that the size of the scale box is 37.5μm (2.5 pixels).

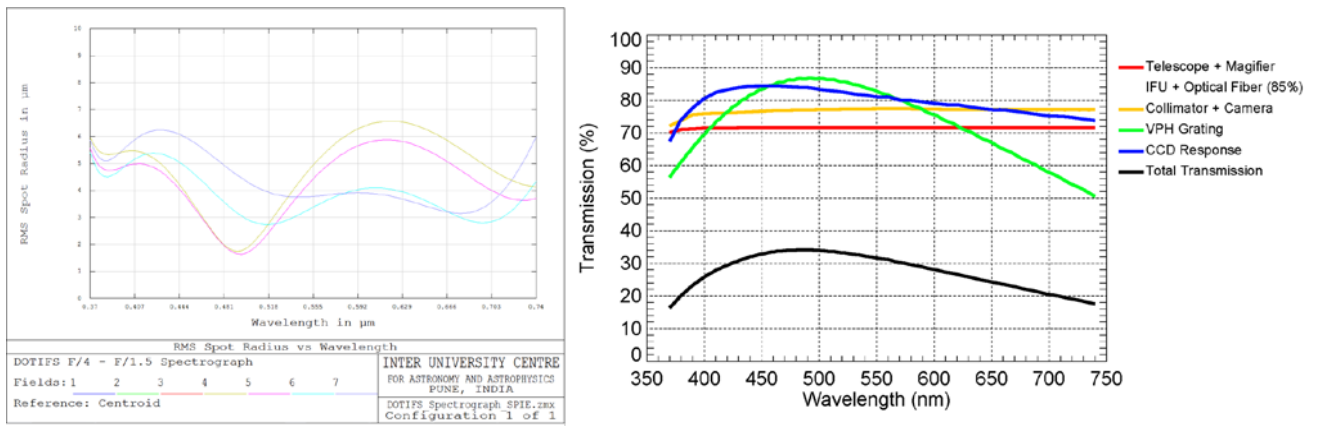

Figure 4. (Left) RMS spot radius vs Wavelength. Y-axis is from 0 to 10μm. (Right) The predicted throughput of the DOTIFS, from the telescope to the CCD. The transmission of the IFU+optical fiber is assumed as 85%, and not shown in the graph, but it is considered in the total transmission.

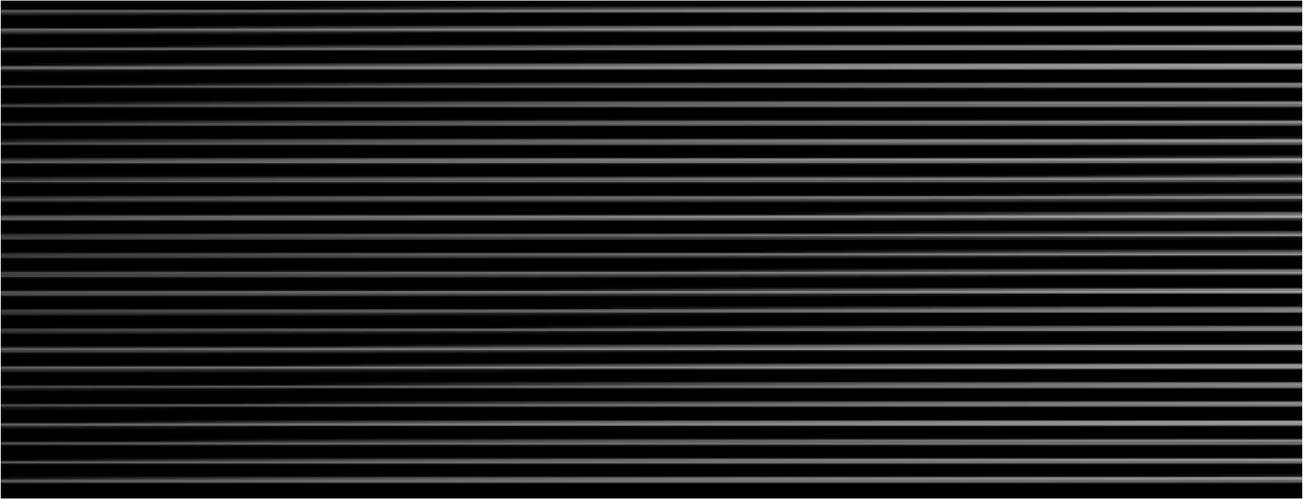

Figure 5. Section of the simulated CCD image of the DOTIFS(flat object). It shows lower left part of the CCD. Each horizontal line represents light from one spatial element.(fiber) A low level of distortion can be found in the image.

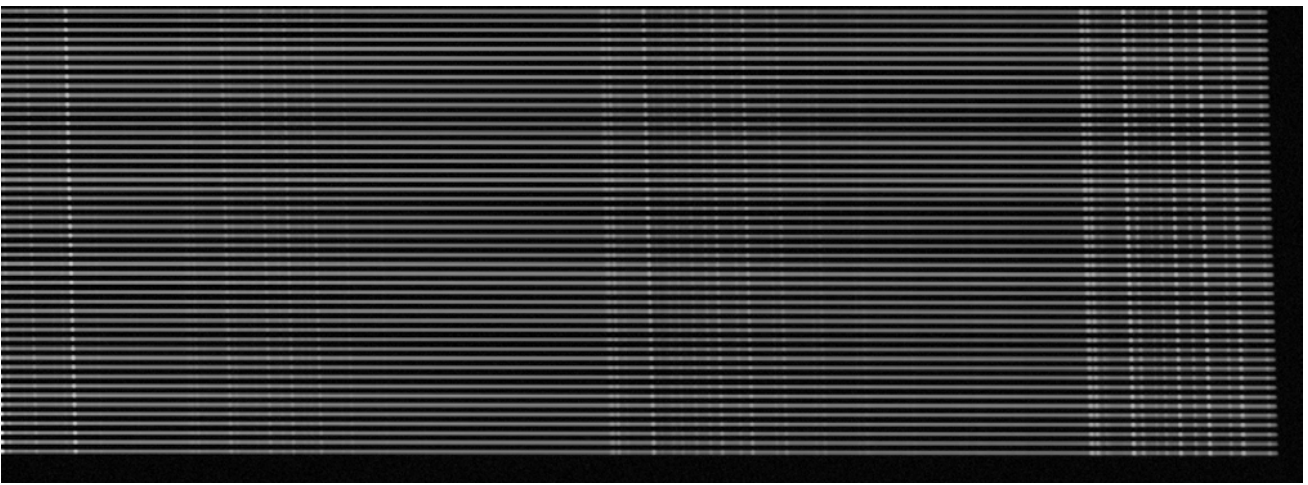

Figure 6. Section of the simulated CCD image of the DOTIFS(artificial object). It shows lower right part of the CCD. Each horizontal line represents light from one spatial element.(fiber) A low level of distortion can be also found in the image. In addition, the bright sky emission lines are noticeable in the image.

## 5. DATA REDUCTION AND SIMULATION SOFTWARE

Data reduction software for DOTIFS has been developed simultaneously with the instrument hardware design. Due to IFS' characteristic, a data reduction pipeline for fiber-fed spectrograph data requires complex reduction procedures. Fortunately, now there are publicly available IFS data reduction tools which can be applied to any IFS, though they need parameter fine tuning for different instruments; eg. P3D[7], or R3D[8]. We plan to use one of them as a starting point, and include some additional features like sky subtraction or treatment of particular observing mode. However, theoretical parameters are not enough to develop the data reduction software due to complex nature of the CCD image of fiber-fed spectrograph. Therefore, it requires actual data or at least similar data to the actual one to develop an optimized software package. The DOTIFS Data Simulator (DDS) has been developed to satisfy that requirement. It produces a simulated CCD image of the spectrograph which includes several practical effect of the actual spectrograph. It is acting like a virtual instrument. It observes virtual object with artificial IFUs, and simulates spectra of two-dimensional object or continuum and emission line sources. Figure 5 and Figure 6 show part of output images of the DDS which are CCD image of a continuum calibration lamp and an artificial object. Based on the result of the DDS, proper data reduction software will be developed before we get the actual observing data. The details of the DDS and status of the data reduction software development will be published separately in future.

## 6. SCHEDULE

The DOTIFS project had been through conceptual design review (October 2012) and baseline design review (April 2013). Progress has been made continuously from then. Currently, some of the major and sample parts have been ordered, arrived, and tested. Part tests for optical fibers including focal ratio degradation, fiber coupling has been made. The optical design for the spectrograph has been finalized and waiting design review. Parts test and order will continue, and optical design for the magnifier will be finalized, soon. Developing deployment system and algorithm is under way. The data simulator has been developed, and the development of exposure time calculator / data reduction software is ongoing. After critical design review, designing all parts and assembling, we expect to commission and see its first light in 2016.